\documentclass[12pt]{mn2e}
\usepackage{graphicx}
\usepackage{epsf}
\usepackage{lscape}
\usepackage{longtable}
\usepackage{rotating}
\usepackage{color}

\newcommand{\aap}{A\&A}

\newcommand{\apj}{ApJ}
\newcommand{\apjl}{ApJ}

\newcommand{\mnras}{MNRAS}
\newcommand{\aj}{AJ}

\newcommand{\pasp}{PASP}

\newcommand{\nat}{Nat}

\begin{document}
\topmargin -0.5in 

\title[Spectroscopy of HUDF-YD3]{VLT/XSHOOTER \& Subaru/MOIRCS Spectroscopy of HUDF-YD3: No Evidence for Lyman-$\alpha$
Emission at $z=8.55$\thanks{Based on observations collected at the European Organisation for Astronomical Research in the Southern Hemisphere, Chile, as part of programme 086.A-0968(B), and based in part on data collected at Subaru Telescope, which is operated by the National Astronomical Observatory of Japan.}}

\author[Andrew J.\ Bunker, et al.\ ]  
{Andrew J. Bunker$^{1}$\thanks{E-mail: a.bunker1@physics.ox.ac.uk}, Joseph Caruana$^{1,2}$, Stephen M. Wilkins$^{1}$, Elizabeth R. Stanway$^{3,4}$, 
\newauthor
Silvio Lorenzoni$^{1}$, 
Mark Lacy$^{5}$,
Matt J.\ Jarvis$^{1,6,7}$ \&  Samantha Hickey$^{6}$\\
$^1$\,University of Oxford, Department of Physics, Denys Wilkinson Building, Keble Road, OX1 3RH, U.K. \\
$^2$\,Leibniz-Institut f\"ur Astrophysik Potsdam, An der Sternwarte 16, 14482 Potsdam, Germany\\
$^{3}$\,H.\ H.\ Wills Physics Laboratory, Tyndall Avenue, Bristol, BS8 1TL, U.K.\\
$^{4}$\,Department of Physics, University of Warwick, Gibbet Hill Road, Coventry, CV4 7AL, U.K.\\
$^{5}$\,NRAO, 520 Edgemont Road, Charlottesville, VA 22903, USA\\
$^{6}$\,Centre for Astrophysics, Science \& Technology Research Institute, University of Hertfordshire, Hatfield, Herts AL10\,9AB, U.K
.\\
$^{7}$\,Physics Department, University of the Western Cape, Cape Town, 7535, South Africa}
\maketitle

\begin{abstract}
We present spectroscopic observations with VLT/XSHOOTER and Subaru/MOIRCS of a relatively
bright $Y$-band drop-out galaxy in the Hubble Ultra Deep Field, first selected by Bunker et al.\ (2010), McLure et al.\ (2010)
and Bouwens et al.\ (2010) to be a likely $z\approx 8-9$ galaxy on the basis of its colours in the {\em HST} ACS and WFC\,3 images.
This galaxy, HUDF.YD3 (also known as UDFy-38135539) has been
targetted for VLT/SINFONI integral field spectroscopy by Lehnert et al.\ (2010), who published a candidate Lyman-$\alpha$
emission line at $z=8.55$ from this source. In our independent spectroscopy using two different infrared spectrographs (5 hours with VLT/XSHOOTER and 11 hours with Subaru/MOIRCS) we are unable to reproduce this line.  We do not
detect any emission line at the spectral and spatial location reported in Lehnert et al.\ (2010), despite the expected signal in
our combined MOIRCS \& XSHOOTER data being $5\,\sigma$. The line emission also seems to be ruled out by the faintness of this object in recently extremely deep F105W ($Y$-band) {\em HST}/WFC\,3 imaging from HUDF12; the line would fall within this filter and such a galaxy should have been detected at $Y_{AB}=28.6$\,mag (\,$\sim 20\sigma$) rather than the marginal $Y_{AB}\approx 30$\,mag observed in the $Y$-band image, $>3$ times fainter than would be expected if the emission lie was real. Hence it appears highly unlikely that the reported Lyman-$\alpha$ line emission at $z>8$  is real, meaning that the highest-redshift sources for which Lyman-$\alpha$ emission has been seen are at $z=6.9-7.2$. It is conceivable that  Lyman-$\alpha$ does not escape galaxies at higher redshifts, where the Gunn-Peterson absorption renders the Universe optically thick to this line. However, deeper spectroscopy on a larger sample of candidate $z>7$ galaxies will be needed to test this.
\end{abstract} 


\section{Introduction}

Candidate galaxies within the first billion years, at redshifts $z>6$, are now being routinely identified through the Lyman-break technique (e.g., Stanway et al.\ 2003; Bunker et al.\ 2004;
Bouwens et al.\ 2006; Hickey et al.\ 2010; McLure et al.\ 2010). Other methods, such as gamma-ray burst follow-up, have also yielded high-redshift galaxies, including
one probably at $z=8.2$ whose spectrum shows a continuum break consistent with Lyman-$\alpha$ (Tanvir et al.\ 2009).
For this growing population of objects with a spectral break consistent with $z>6$, proper spectroscopic
confirmation is important, rather than broad-band photometry or putative breaks in low S/N spectroscopy.

The main feature which might be detectable is Lyman-$\alpha$ emission, resulting from photoionization of H{\scriptsize ~II}
regions by star formation. However, the discovery of the Gunn-Peterson complete absorption trough below Lyman-$\alpha$ (Gunn \& Peterson 1965, Scheuer 1965) in SDSS and UKIDSS QSOs at
redshifts beyond $z\approx 6.2$ (Becker et al.\ 2001; Fan et al.\ 2001, 2006; Mortlock et al.\ 2011) shows that the Universe is on average
optically thick to this line at earlier times. This suggests that $z\approx 6$ lies at the end of the Epoch of Reionization, whose mid-point may have occurred at $z\approx 11$, according to results from WMAP (Dunkley et al. 2009). It has been speculated that a large enough H{\scriptsize ~II} bubble around
a galaxy might render this line non-resonant when it encounters the neutral IGM, so Lyman-$\alpha$ might possibly
emerge after all even during the Gunn-Peterson era. While spectroscopy has confirmed $i'-$drop Lyman-break
galaxies at $z\approx 6$ (e.g. Bunker et al.\ 2003; Stanway et al.\ 2004; Stanway et al.\ 2007),
 spectroscopic follow-up of $z>6$ sources
has had mixed success. Vanzella et al.\ (2011) showing convincing line emission from two Lyman-break
galaxies at $z=7.0-7.1$, and with one marginal $z>6.4$ emission line out of 17 targets reported by Stark et al.\ (2010)
and another marginal detection (out of 7 targets) from Fontana et al.\ (2010; see also Pentericci et al.\ 2011). More recently,
Schenker et al.\ (2012) targetted 19 Lyman-break galaxies with photometric redshifts $6.3<z<8.8$, but found
only one object at $z>7$ ($z=7.045$ with another more marginal candidate at $z=6.905$). A similar survey
by Caruana et al.\ (2012) failed to detect Lyman-$\alpha$ from any $z$-band or $Y$-band drop-outs at
$z>7$. Another way to isolate emission lines directly is
narrow-band imaging, and Suprime-Cam on  the Subaru telescope has revealed a Lyman-alpha emitter spectroscopically confirmed
to be at $z=6.96$ (Iye et al.\ 2006) with another three possible candidates (Ota et al.\ 2010),The $z=6.96$ source
was subsequently observed by Ono et al.\ (2012), who also confirmed two $z$-drop galaxies a $z=6.844$ and $z=7.213$
with Lyman-$\alpha$ emission. Another  narrow-band-selected Lyman-$\alpha$ emitter has recently been confirmed at $z=7.215$ (Shibuya et al.\ 2012).

There
has only been one recent claim of line emission beyond $z\approx 7.2$, despite the large number of Lyman-break
candidates now known at these redshifts. Lehnert et al.\ (2010) presented a VLT/SINFONI spectrum of one of the brightest
$Y$-drops in the WFC\,3 imaging of the Hubble Ultra Deep Field, which had previously been independently
selected on the basis of its broad-band ACS and WFC\,3 photometry by three independent groups
(the galaxy HUDF-YD3
in the catalogue of
Bunker et al.\ 2010, object 1721 in McLure et al.\ 2010\footnote{The naming of this galaxy changes to HUDF\_2003 in McLure et al.\ (2011).}, and galaxy UDFy-38135539 in Bouwens et al.\ 2010\footnote{We note that in a subsequent paper (Bouwens et al.\ 2011), this galaxy has a different identification number, UDFy-38125539.}). The Lehnert et al.\ (2010)
spectrum shows a $6\,\sigma$ line at 11616\,\AA\ which is consistent with being Lyman-$\alpha$ emission at $z=8.55$, close
to the photometric redshift of $z=8.45$ from McLure et al.\ (2010). If real, the emergence of Lyman-$\alpha$ emission well within the Gunn-Peterson epoch would
have significant implications for the size of H{\scriptsize ~II} regions around galaxies, and would mean that Lyman-$\alpha$ might still be
a useful redshift indicator for very distant galaxies even at a time when most of the Universe is optically thick to this line.
However, previous claims of Lyman-$\alpha$ emission at similarly large redshift (e.g. Pell\'{o} et al.\ 2004; Chen, Lanzetta \& Pascarelle 1999) have not survived critical
re-analysis (e.g. Bremer et al.\ 2004; Weatherley, Warren \& Babbedge 2004; Stern et al.\ 2000). In this paper we re-observe the galaxy
HUDF.YD3 from Bunker et al.\ (2010) with VLT/XSHOOTER  and Subaru/MOIRCS spectroscopy to see if we can repeat
the detection of Lyman-$\alpha$ at $z=8.55$ made by Lehnert et al.\ (2010).

The structure of this paper is as follows. We describe our spectroscopic observations in Section~\ref{sec:obs},
and present the results of the spectroscopy and constraints from the {\em HST} imaging in Section~\ref{sec:discuss}.
Our conclusions are given in Section~\ref{sec:concs}, and throughout we adopt a standard $\Lambda$CDM cosmology
with $\Omega_M=0.3$, $\Omega_{\Lambda}=0.7$ and $H_0=70\,{\rm km\,s^{-1}\,Mpc^{-1}}$. All magnitudes are on the AB system.

\section{Observations and Data Reduction}
\label{sec:obs}

\subsection{Observations with VLT/XSHOOTER}

We observed the $Y$-band drop-out high-redshift galaxy candidate HUDF-YD3 using
the XSHOOTER spectrograph (D'Odorico et al.\ 2006) on the ESO VLT-UT2 (Kueyen) as part of programme 086.A-0968(B) (PI: A.\ Bunker).
XSHOOTER is an echelle
spectrograph, with UV, visible and near-infrared channels obtaining near-continuous
spectroscopy from 0.3\,$\mu$m to 2.48\,$\mu$m. We will focus here on the near-infrared spectroscopy
around 1.12\,$\mu$m, at the location of the emission line claimed by Lehnert et al.\ (2010) in
their ESO/VLT SINFONI spectroscopy.

The main target, HUDF.YD3, has a position RA=03:32:38.135, Dec.= $-$27:45:54.03 (J2000), 
with coordinates from Lorenzoni et al.\ (2011). We set the position angle of the $11''$-long XSHOOTER
slit to 54.0\,degrees East of North.
We set the central coordinates to be RA=03:32:38.086, Dec.=$-$27:45:54.71 (J2000), such that
HUDF.YD3 lay $1''$ away along the slit long axis. We dithered the observations in an ABBA
sequence at positions $+3''$ and $-3''$ from the central coordinates along the slit long axis (i.e. a `chop' size of $6''$),
so that the expected position of HUDF.YD3 should be $+4''$ above the slit centre in the `A'
position, and $-2''$ in the `B' position. 
To acquire the target, we first peaked-up on a bright star $76.1''$ East and $10.6''$ South of the
desired central pointing, then did a blind offset. ESO guarantee an accuracy of $<0\farcs1$ for
an offset of this size, providing the guide star remains the same (which was the case), meaning
that the positional uncertainty is less than 10 per cent of
the slit width used (1\farcs2) -- we note that our blind offset of $1\farcm3$ is less than that of $1\farcm5$ used
by Lehnert et al.\ (2010). The XSHOOTER slit width is also much greater than the limit of $<0\farcs4$ set on any positional offset between the continuum
position and that of the claimed line emission from Lehnert et al.\ (2010).

The XSHOOTER observations were conducted in 6 observing blocks, each of 1\,hour duration (49\,min of
which was on-source)
and consisting of a single ABBA sequence with three exposures of the near-IR arm of duration
245\,s at each A or B position. The observations were taken on the nights of UT 2010 December 27, 29, 30 \& 31,
with two observing blocks taken on the nights of UT 2010 December 29 \& 30 and single observing blocks
on the other nights. Observing conditions were reported to be clear, and the seeing conditions were typically $0\farcs5-0\farcs6$ FWHM
(from DIMM measurements taken at the time, and we checked this was consistent with observations of standard stars taken
close in time to our observations). One of the two observing blocks taken on UT 2010 December 30 had significantly worse seeing of $1\farcs2$ FWHM,
and we reduced the full dataset twice, with and without this bad-seeing block. This did not appear to have a significant impact on the final results.
Our total on-source exposure time for HUDF.YD3 with XSHOOTER was 4.9\,hours, with 4.1\,hours taken in good seeing
conditions of $0\farcs5-0\farcs6$.

From unblended spectral lines in the calibration arc lamp spectra and in the sky spectra we measured a
spectral resolution of $R=\lambda/\Delta \lambda_{\rm FWHM}=5000$. 
We note that our arc and sky lines
fill the slit, so for compact sources which do not fill the slit in the good seeing
the resolution will be better than this (we expect this to be the case for HUDF.YD3).

We reduced the XSHOOTER spectroscopy in two different ways. We initially used the
ESO pipeline (Modigliani et al.\ 2010), which used the two-dimensional arc spectra
through a pinhole mask to rectify the spectra both spatially and spectrally (the echelle
spectra exhibited significant spatial curvature and a non-linear wavelength scale),
mapping on to a final output spectral scale of 1\,\AA\,pix$^{-1}$ (from an original scale
of about 0.5\,\AA\,pix$^{-1}$ at wavelengths close to 11616\,\AA ), and a
spatial scale of 0\farcs21 (from an original scale of 0\farcs24). The pipeline applied
a flat-field, identified and masked cosmic ray strikes using the algorithm of
van Dokkum (2001), differenced the two dither positions to remove the sky
to first order, and combined the different echelle orders together into
a continuous spectrum (taking into account the different throughputs in
different overlapping echelle orders) before 
spatially registering and combining the data taken at the two dither positions,
and removing any residual sky background.

We note that the ESO pipeline interpolates the data onto
a uniform grid, which has the effect of correlating the noise (making the measured
noise an underestimate of the true noise), and also potentially spreading the effect
of cosmic ray strikes and hot pixels around neighbouring pixels. Hence, we also did
our own independent reduction of the XSHOOTER spectroscopy, where we did
not interpolate the data, keeping each pixel statistically independent. The data were
flat-fielded using halogen lamp spectra (that had been normalized by division by
the spectral shape of the lamp), and multiple exposures at each dither position were averaged using
the IRAF task {\tt imcombine}, using a Poisson noise model to reject cosmic ray strikes.
The two dither positions were then combined,
with known hot pixels masked. 
The measured noise in the reduced two-dimensional spectrum was close to
the expected Poisson noise from the sky background, dark current and readout noise,
and when combining several exposures the noise (normalized to unit time) decreased as $\sqrt{\rm time}$ as
expected.
The wavelength and spatial position of each pixel in
the two-dimensional spectrum was determined from the sky lines in the actual data
and the arc line calibration spectra
taken through a pinhole mask. The spectrograph setup seemed very stable between different nights
of observation, with shifts of only $\approx 0.2$ pixels between nights.
Residual skyline emission was removed using the {\tt background}
task in IRAF. The expected position of Lyman-$\alpha$ at $z=8.55$ appears at the red end of order 23 
(and at the blue end of order 22, but the
throughput here is lower). The pipeline optimally combines the orders
of the echelle spectrum, but in our independent reduction we inspected both echelle orders
separately. The depths quoted in Section~\ref{sec:discuss} come from the deepest
spectrum, order 22. 

We obtained a flux calibration from observations of spectrophotometric
standard stars taken over UT 2010 December 26--31, around the dates when our HUDF-YD3
spectra were obtained. We base our flux calibration on observations of the standard star LTT\,3218 on UT 2010 December 28
taken in good seeing of  $0\farcs6$, which is a close match to the seeing in our spectroscopy of HUDF.YD3. We have checked the
shape of the spectral response is similar on other nights where the flux standards LTT\,3218
and Feige 110 were taken in worse seeing. We note that although the region of interest around 11616\,\AA\
is close to atmospheric absorption features, the depth of the absorption at this wavelength was not great
and was stable night-to-night. Around our wavelength of interest, 1 count in a single 245\,s integration
corresponds to a line flux of  $3.4\times 10^{-19}\,{\rm erg\,cm^{-2}\,s^{-1}}$.

\subsection{Observations with Subaru/MOIRCS}

We observed the HUDF with slitmask spectroscopy in the near-infrared using
the MOIRCS instrument (Suzuki et al.\ 2008; Ichikawa et al.\ 2006) on Subaru. MOIRCS was used in
slitmask mode, which uses two detectors with a combined field of view of $7'\times4'$, although
there is vignetting beyond a diameter of $6'$ from the field centre. Unfortunately a filter wheel
issue meant that one of the two detectors was unusable, so we ensured
that all our priority HUDF targets were placed in the other half of the slitmask.
Accurate alignment of the slitmask was achieved by centering 5 stars
within 3\farcs5-wide boxes to an accuracy of $\approx$0\farcs1.
One of the slits was used
to target HUDF.YD3, and this slit was 4\farcs5 in length, with the long
axis of the slit (the Position Angle of the mask) set to +57degrees East
of North. We observed the mask
with individual integrations of 1200\,s, moving the telescope along the slit
axis by a small dither size of 2\farcs0--2\farcs5 in an ABABAB sequence
to enable background subtraction.
We observed the HUDF mask on U.T.\ 2010 October 21 \& 22, with a slit
width of 1\farcs0, and using the zJ500 grism. This instrument set-up
has a spatial scale of 0\farcs117\,pix$^{-1}$ and a spectral scale
of 5.57\,\AA\,pix$^{-1}$. The resolving power for objects which fill the slit is 
$R=\lambda/\Delta\lambda_{FWHM}=300$ (determined from Thorium-Argon
arc lines),
but the typical seeing was 0\farcs5 FWHM so for unresolved sources (such
at most of the high-redshift galaxies targetted) the resolving power is $R=500$.
On U.T.\ 2010 October 21 we obtained 8 exposures of 1200\,s, with a dither
step of 2\farcs5 (i.e.\ placing the target at $+$1\farcs25 and $-$1\farcs25
above and below the slit centre). On U.T.\ 2010 October 22 we reduced the dither step to 2\farcs0,
given the good seeing, and obtained another 12 exposures of 1200\,s
for a total integration time of 400\,min (6.67\,hours) in October 2010.
We observed the same slitmask targets again with Subaru/MOIRCS on U.T.\ 2010 December 07,
obtaining 12 exposures of 1200\,s (a total of 4\,hours) with a dither size of 2\farcs0.
To take full advantage of the good seeing at Subaru (which again was 0\farcs5 for the December 2010
observations) we used a new mask design with the same objects targetted but with
the slit width reduced to 0\farcs7, instead of 1\farcs0 used in October 2010,
achieving a resolving power $R=500$. The narrower slits reduced
the sky background, while still capturing most of the flux from the unresolved galaxies,
significantly improving our sensitivity at the expected Lyman-$\alpha$ wavelength, 11616\,\AA\ (which is close
to OH sky lines).

We reduced the MOIRCS data using standard techniques in {\tt IRAF}, treating
the October 2010 and December 2010 separately due to the different slit widths. The average of many dark currents was 
subtracted from each frame, and a flat field applied (obtained from dome flats, normalized by
the spectrum of the lamp). We then combined separately all the data frames in the A position of the
dither, using {\tt ccdclip} in {\tt imcombine} to reject cosmic rays given the parameters of the detector
(gain of 3.3\,$e^-$\,count$^{-1}$ and readout noise of 29\,$e^-$\,pix$^{-1}$). The same was done for the B positions,
and the combined B frame was subtracted from the combined A frame to remove the sky background to first order. This
resulted in a frame where we expect a positive signal from a source at position A, and a negative signal at position
B (offset along the slit by the dither step). We then shifted and combined these signals, and residual sky emission
was subtracted through polynomial fits along the slit length.

Flux calibration was achieved through observation of the A0 star HIP116886, and checked against
the flux of the alignment stars of known $J$-band magnitude seen through the 3\farcs5-wide alignment boxes in the data frames.
Around 11616\,\AA\ (the wavelength of interest), 1 count in an individual 1200\,s exposure
corresponds to a line flux of $5.6\times10^{-20}\,{\rm erg\,cm^{-2}\,s^{-1}}$.

\section{Discussion}
\label{sec:discuss}
\subsection{Upper Limits on the Lyman-$\alpha$ Flux at $z=8.55$ from VLT/XSHOOTER}

We measure the observed flux at the location of HUDF.YD3 in the XSHOOTER
long-slit spectrum, 1\,arcsec
above (North-East of) the slit centre, and at the expected wavelength of Lyman-$\alpha$
from Lehnert et al.\ (2010), $\lambda_{vac}=11615.6$\,\AA\ ($\lambda_{air}=11612.4$\,\AA ).
We detect no sign of an emission at this location. We perform spectrophotometry
using a square aperture, of extent 5\,\AA\ (10 pixels across in the wavelength domain
for our own reduction of the data, and 5 pixels in the pipeline reduction), which is more
than twice as large as the width of a spectrally unresolved line. For the spatial
extent of our aperture, we adopt 3 pixels (0\farcs72) for our reduction, and 4 pixels
(0\farcs84) for the pipeline reduction (the XSHOOTER pipeline resamples the original
pixel scale slightly), which is marginally larger than the size of the seeing disk. 
Hence in our reduction, where the pixels are unresampled, we
measure the flux total flux in 30 independent pixels, and from the pipeline data (which
involves interpolation) the flux is measured over 20 pixels.

\begin{figure}
 \begin{center}
   \resizebox{0.47\textwidth}{!}{\includegraphics{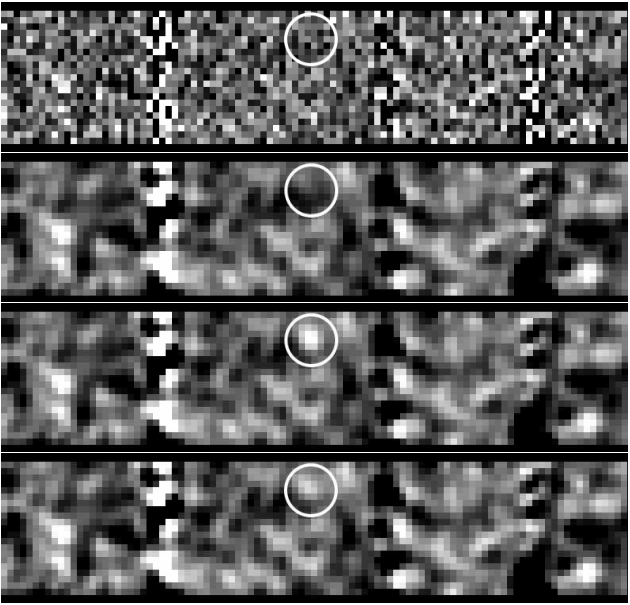}} 
 \end{center}
 \caption{The pipeline-calibrated XSHOOTER spectrum, with the location
 of HUDF.YD3 (1arcsec above the centre of the long slit) and the expected
 wavelength of the Lyman-$\alpha$ emission reported by Lehnert et al.\ (2010)
 marked with a white circle. Wavelength increases from left to right, and 
 we show the 50\,\AA\ either side of 11616\,\AA , and the vertical axis is
 the 4.4\,arcsec covered in both nod positions of the XSHOOTER slit.
 From top to bottom: (a) the pipeline-reduced data; (b) the pipeline-reduced data
 convolved with a Gaussian of $\sigma=1$\,pixel (1\,\AA , 0\farcs21).
 (c) a fake source with the same line flux ($6\times 10^{-18}\,{\rm erg\,cm^{-2}\,s^{-1}}$)
 and wavelength as the Lehnert et al.\ (2010) line added into the frame. We assume
 a spatially and spectrally unresolved source, with FWHM$=0\farcs6$ spatially
 and FWHM$=2.3$\,\AA\ spectrally. The resulting frame has been smoothed with a Gaussian
 with $\sigma=1$\,pixel. (d) a fake source with the same line flux and wavelength as the 
 Lehnert et al.\ (2010) line added into the frame, with a broader FWHM$=5$\,\AA\ and again
 unresolved spatially. The three vertical lines of higher noise are due to night sky emission lines.}
 \label{fig:2Dspec}
\end{figure}

We detect no significant line emission -- we measure the flux in our aperture to be $(-0.45\pm1.2)\times10^{-18}\,{\rm ergs\,cm^{-2}\,s^{-1}}$,
where the error is the measured  $1\,\sigma$ noise. We also move the aperture
by $\pm2$\,pixels in $x$ and $y$ in a $3\times 3$ grid to bracket the maximum  uncertainty
in the position of the Lehnert et al.\ (2010) Lyman-$\alpha$ emission ($<0\farcs4$),
and we have no detection of line emission at any of these locations.
Our measured noise is consistent
with the online ESO Exposure Time Calculator for XSHOOTER. We note that  
the Lehnert et al.\ (2010) line flux would be detected at the $5\,\sigma$ level if
all the line emission fell within our aperture. 
In order to quantify the expected flux, corrected for aperture losses, we created
artificial emission lines to add in at this spatial and spectral location, as shown in
Figure~\ref{fig:2Dspec}. From the {\em HST}/WFC3
imaging, HUDF.YD3 should be unresolved in our 0\farcs5-0\farcs6 FWHM seeing. While
it is conceivable that resonantly-scattered Lyman-$\alpha$ line emission may come from a larger halo
than the stellar UV continuum (e.g., Bunker, Moustakas \& Davis 2000; Steidel et al.\ 2011), the
emission line reported in HUDF.YD3 by Lehnert et al.\ (2010) is compact spatially (unresolved
in their 0\farcs6 seeing). Hence we adopt a Gaussian profile for the spatial extent with a FWHM of 0\farcs6.
For the spectral direction, we also adopt a Gaussian profile for the fake sources, and consider
two scenarios for the velocity width. We note that the emission line in Lehnert et al.\ (2010) is
unresolved or marginally resolved (with a FWHM of 9.2\,\AA , only $1\,\sigma$ larger than the resolution
of SINFONI which has $R=1580$). Our first scenario has the source spectrally unresolved by XSHOOTER,
which has a higher resolving power of $R=5000$ (so $\Delta\lambda_{FWHM}=2.3$\,\AA ). In this case, our 
photometric aperture would capture 87 per cent of the line flux, and we would expect
a line with the same total flux as in the Lehnert et al.\ (2010) to be detected at $4.5\,\sigma$. The
second scenario takes the reported (marginally-resolved) spectral width of 9.2\,\AA\ from Lehnert et al.\ (2010),
deconvolves this with the SINFONI resolution to obtain an intrinsic line width of 5.5\,\AA\ FWHM (140\,km\,s$^{-1}$),
then convolve this with our spectral resolution for XSHOOTER to obtain an observed line width of 6\,\AA\ FWHM.
For this broader line, our photometric aperture captures 66 per cent of the line flux, and we would expect
a line with the same total flux as in the Lehnert et al.\ (2010) to be detected at $3.5\,\sigma$.
Our XSHOOTER spectroscopy appears to rule out the existence of the Lyman-$\alpha$ line reported
by Lehnert et al.\ (2010) at the $3.5-4.5\,\sigma$ level, depending on the velocity width of the line.

\subsection{Upper Limits on the Lyman-$\alpha$ Flux at $z=8.55$ from Subaru/MOIRCS}

For the Subaru/MOIRCS data, we used an aperture of size $5\times 5$\,pixels centred
on the expected position of Lyman-$\alpha$, corresponding to $0\farcs6\times 28$\,\AA ,
which is slightly larger than a resolution element. The $1\,\sigma$ noise within this aperture was measured
to be $2.1\times 10^{-18}\,{\rm erg\,cm^{-2}\,s^{-1}}$ for the December 2010 observations,
and $2.4\times 10^{-18}\,{\rm erg\,cm^{-2}\,s^{-1}}$ for the October 2010 observations (which
had higher noise due to the wider slit used and hence more sky emission).
For the 0\farcs5 seeing and a spectrally
unresolved line (where the resolution is 600\,km\,s$^{-1}$), such an aperture encloses
68 per cent of the total flux. Hence we would expect an emission line of the flux
and wavelength reported by Lehnert et al.\ (2010) to be present at the $2.7\,\sigma$ level
in our total Subaru/MOIRCS spectrum, with most of the sensitivity coming from the
December 2010 data using a narrower slit (where such a line should be present at the $2.0\,\sigma$
level). However, in both sets of MOIRCS observations this line is undetected, with a total
aperture-corrected flux of $(1.6\pm 3.1)\times 10^{-18}\,\rm {erg\,cm^{-2}\,s^{-1}}$ for the deeper December 2010,
and a total flux of $(-0.1\pm 2.3)\times 10^{-18}\,{\rm erg\,cm^{-2}\,s^{-1}}$ when combining
all the MOIRCS observations together from all three nights (using inverse-variance weighting).
 Although the MOIRCS spectrum
is less deep than our XSHOOTER spectrum (on account of the lower spectral resolution
of MOIRCS), the MOIRCS spectrum still is useful because we are very confident of slit
position, at the $0\farcs1$ level, due to the number of alignment stars used to position the
slitmask. For the XSHOOTER spectrum (and indeed the Lehnert et al.\ SINFONI spectrum)
a blind offset was performed from a nearby star, which does introduce some uncertainty --
although the tolerance is supposed to be less than 0\farcs4 (the maximum positional
uncertainty for the Lyman-$\alpha$ line given by Lehnert et al.\ 2010).
Both our VLT/XSHOOTER and Subaru/MOIRCS spectroscopy yield consistent results:
we see no emission line at $\lambda_{vac}=11615.6$\,\AA\ at the position of HUDF.YD3,
whereas if the flux reported by Lehnert et al.\ (2010) is accurate we should have seen
a signal at $3.5-4.5\,\sigma$ with XSHOOTER and $2.7\,\sigma$ with MOIRCS.
Combining the results from two different spectrographs with inverse-variance weighting, the Lehnert
et al.\ (2010) line flux is ruled out at the $5\,\sigma$ level.

\subsection{{\em HST} Photometry}

Our VLT/XSHOOTER and Subaru/MOIRCS spectroscopy of HUDF.YD3 strongly suggests that
there is no line at the wavelength and line flux claimed by Lehnert et al.\ (2010) on the basis
of their VLT/SINFONI spectroscopy. We now briefly consider whether the Lehnert et al.\ (2010)
emission line would have been consistent with the {\em HST}/WFC\,3 broad-band photometry
of this object reported by several groups (Bunker et al.\ 2010; Bouwens et al.\ 2010; McLure
et al.\ 2010; Lorenzoni et al.\ 2011). The first WFC3 observations of the HUDF taken as part
of the programme GO-11563 (HUDF09, PI: G.\ Illingworth) used the F105W (``$Y$-band"), F125W (``$J$-band")
and F160W (``$H$-band") filters. An emission line at 11615.6\AA\ would lie entirely within
the $Y$-band (and also within the wide $J$-band), in the area of peak transmission
of the sharp-sided $Y$-filter. If we take the line flux of $6.1\times 10^{-18}\,{\rm erg\,cm^{-2}\,s^{-1}}$,
then this would be equivalent to an observed broad-band magnitude of $Y_{AB}=28.89$. There
should also be a contribution from the UV-continuum photons long-ward of Lyman-$\alpha$ (assuming
near-total absorption by the Lyman-$\alpha$ forest at shorter wavelengths). Only 20 per cent of
the $Y$-band filter transmission would lie at wavelengths above Lyman-$\alpha$ at the claimed redshift of $z=8.55$
(Lehnert et al.\ 2010), which would imply a broad-band magnitude from the claimed line and continuum
of $Y_{AB}=28.57$. In calculating the UV flux density we use the measured {\em HST}/WFC3
broad-band magnitudes of $J_{AB}=28.18\pm0.13$ and $H_{AB}=28.10\pm0.13$ (Lorenzoni et al.\ 2011), and adopt
a rest-UV spectral slope of $f_{\lambda}\propto\lambda^{-2.0}$, consistent with the {\em HST}/WFC3 colours after we
correct the $J$-band for the small fraction of flux within this filter that would fall below Lyman-$\alpha$ (a correction
of 0.15\,mag, comparable to the measurement error on the magnitudes).
We note that  HUDF.YD3 has a magnitude fainter
than the  2\,$\sigma$ limiting magnitude of $Y_{AB}(2\,\sigma)=29.65$ in a 0\farcs6-diameter aperture for the HUDF09 data,
and is formally undetected in the HUDF09 WFC3 imaging (Bunker et al.\ 2010; McLure et al.\ 2010; Bouwens et al.\ 2010; Lorenzoni et al.\ 2011). 

The first WFC3 imaging with the F105W filter was 14 orbits (with another 4 orbits compromised by cosmic ray persistence),
and since then this field has been extensively targetted for further imaging with this filter as part of the HUDF12
programme (Ellis et al.\ 2013) increasing the depth to 100 orbits in total.
In this deeper data, McLure et al.\ (2013) and Schenker et al.\ (2013) report a faint detection of a corresponding
object (labelled UDF12-3813-5540 in their catalogues)
of $Y_{AB}=30.1\pm 0.2$, close to the $5\,\sigma$ limit (using an aperture of 0\farcs4 diameter, although apparently they have not applied
an aperture correction to the $\approx$70 per cent of flux enclosed, so the total magnitude will be $\approx 0.3$mag
brighter, $Y_{AB}=29.8$). This
is a factor of $>3$ times fainter than the expected magnitude of $Y_{AB}=28.57$ if the emission line flux reported
by Lehnert et al.\ (2010) was real and due to Lyman-$\alpha$ from a Lyman-break galaxy at $z=8.55$.

 Hence the broad-band photometry in the $Y$-band is inconsistent with the Lehnert et al.\ (2010)
line flux and redshift being real -- if the line was real, then the deep HUDF12 {\em HST}/WFC3 $Y$-band should have
obtained a clear $15\,\sigma$ detection, whereas the actual result was close to the  $5\,\sigma$
limiting magnitude. The broad-band photometry alone seems to rule out the claimed line flux from Lehnert et al.\ (2010)
at high significance. Consistency with the Lehnert et al.\ result would require both that the broadband flux is 
greatly underestimated due to noise and that the line flux is overestimated, a coincidence which is statistically unlikely.

\section{Conclusions}
\label{sec:concs}

We have presented spectroscopic observations with VLT/XSHOOTER and Subaru/MOIRCS of a relatively
bright $Y$-band drop-out galaxy in the Hubble Ultra Deep Field, first selected by Bunker et al.\ (2010), McLure et al.\ (2010)
and Oesch et al.\ (2010) to be a likely $z\approx 8-9$ galaxy on the basis of its colours in the {\em HST} ACS and WFC\,3 images.
This galaxy, HUDF.YD3 (from the catalogues of Bunker et al.\ 2010 and Lorenzoni et al.\ 2011) has been
targetted for VLT/SINFONI integral field spectroscopy by Lehnert et al.\ (2010), who published a candidate Lyman-$\alpha$
emission line at $z=8.55$ from this source. In our independent spectroscopy using two different infrared spectrographs we are unable to reproduce this line. In our 5\,hour spectrum with XSHOOTER with a moderately high resolving power of $R=5000$, the line flux of $6.1\times 10^{-18}\,{\rm erg\,cm^{-2}\,s^{-1}}$ reported by Lehnert et al.\ (2010) should have resulted in a detection at the $3.5-4.5\,\sigma$ level (depending on the
velocity width of the line), and in our low-resolution ($R=500$) 10.7\,hour MOIRCS spectrum this line flux would correspond to a $2.7\,\sigma$ signal. We do not
detect any emission line at the spectral and spatial location reported in Lehnert et al.\ (2010), despite the expected signal in
our combined MOIRCS \& XSHOOTER data being $5\,\sigma$. The line emission also seems to be ruled out by the faintness of this object in the very deep $Y$-band {\em HST}/WFC\,3 image ($Y_{AB}=30.1$); the line would fall within this filter, and the corresponding magnitude of $Y_{AB}=28.57$ should have been detected at $\approx 20\,\sigma$ rather than the marginal $5\,\sigma$ observed. Hence it appears highly unlikely that the reported Lyman-$\alpha$ line emission at $z>8$  is real, meaning that the highest-redshift sources for which Lyman-$\alpha$ emission has been seen are at $z=6.96-7.2$. It is conceivable that  Lyman-$\alpha$ does not escape galaxies at higher redshifts, where the Gunn-Peterson absorption renders the Universe optically thick to this line. However, deeper spectroscopy on a larger sample of candidate $z>7$ galaxies will be needed to test this.

\subsection*{Acknowledgements}

We gratefully acknowledge Chris Willott, Daniel Stark and the NIRSpec Instrument Science Team for useful discussion,
and thank Richard Ellis, Matt Lehnert and the referee (Masanori Iye) for useful comments on this manuscript.
We are indebted to Ichi Tanaka of the Subaru Observatory, NAOJ, for his invaluable assistance in
designing the MOIRCS slitmask and during the observations.
Based on observations made with the NASA/ESA {\em Hubble} Space Telescope associated with programmes \#GO-11563 \& \#GO-12498,
obtained from the Data Archive at the Space Telescope Science Institute, which is operated by the Association
of Universities for Research in Astronomy, Inc., under NASA contract
NAS 5-26555. 
MJJ acknowledges the support of a RCUK
fellowship. 
JC and SL are supported by the Marie Curie Initial Training Network ELIXIR of the European Commission under
contract PITN-GA-2008-214227.

\bsp

\end{document}